\documentclass[reprint,twocolumn,superscriptaddress,nofootinbib,aps,prd]{revtex4-1}
\usepackage{lipsum,url}
\usepackage{marvosym}
\usepackage{graphicx,pslatex}
\usepackage{amssymb}
\usepackage{bbold}
\usepackage{color}
\usepackage{tikz}
\usetikzlibrary{plotmarks}
\newcommand\marksymbol[2]{\tikz[#2,scale=0.85]\pgfuseplotmark{#1};}
\usepackage{amsmath}
\def\spose#1{\hbox to 0pt{#1\hss}}
\def\ltapprox{\mathrel{\spose{\lower 3pt\hbox{$\mathchar"218$}}
 \raise 2.0pt\hbox{$\mathchar"13C$}}}
\def\gtapprox{\mathrel{\spose{\lower 3pt\hbox{$\mathchar"218$}}
 \raise 2.0pt\hbox{$\mathchar"13E$}}}

\DeclareMathOperator{\Tr}{Tr}
\newcommand{\1}{1\!\!\!\bot}

\allowdisplaybreaks

\begin{document}


\title{{\Large {\bf
       Faddeev-Popov Matrix in Linear Covariant Gauge:
       First Results}}}

\author{Attilio~Cucchieri}
\email{attilio@ifsc.usp.br}
\affiliation{Instituto de F\'\i sica de S\~ao Carlos, Universidade de S\~ao Paulo,
             C.P.\ 369, 13560-970 S\~ao Carlos, SP, Brazil}

\author{David~Dudal}
\email{david.dudal@kuleuven.be}
\affiliation{KU Leuven Kulak, Department of Physics,
            Etienne Sabbelaan 53 bus 7657, 8500 Kortrijk, Belgium}
\affiliation{Ghent University, Department of Physics and Astronomy,
             Krijgslaan 281-S9, 9000 Gent, Belgium}

\author{Tereza~Mendes}
\email{mendes@ifsc.usp.br}
\affiliation{Instituto de F\'\i sica de S\~ao Carlos, Universidade de S\~ao Paulo,
             C.P.\ 369, 13560-970 S\~ao Carlos, SP, Brazil}

\author{Orlando~Oliveira}
\email{orlando@fis.uc.pt}
\affiliation{CFisUC, 
             Departamento de F\'\i sica, Universidade de Coimbra,
             3004-516 Coimbra, Portugal}

\author{Martin~Roelfs}
\email{martin.roelfs@kuleuven.be}
\affiliation{KU Leuven Kulak, Department of Physics,
            Etienne Sabbelaan 53 bus 7657, 8500 Kortrijk, Belgium}

\author{Paulo~J.~Silva}
\email{psilva@teor.fis.uc.pt}
\affiliation{CFisUC, 
             Departamento de F\'\i sica, Universidade de Coimbra,
             3004-516 Coimbra, Portugal}

\begin{abstract}
We discuss a possible definition of the Faddeev-Popov
matrix for the minimal linear covariant gauge on the
lattice and present first results for the ghost
propagator.
We consider Yang-Mills theory in four space-time dimensions, for SU(2)
and SU(3) gauge groups.
\end{abstract}

\maketitle


\section{Introduction}

The behavior of Green's functions in the infrared limit
of Yang-Mills theory has been studied extensively in Landau gauge, both
analytically and numerically \cite{Alkofer:2000wg,Landau1,Landau2}.
However, since the evaluation of propagators and vertices depends on
the gauge condition, a natural extension of these works
would be to consider the linear covariant gauge (LCG),
which depends on a gauge-fixing parameter $\xi$ and has
the Landau gauge as a limiting case, corresponding to $\xi = 0$.
On the lattice, there have been a few studies
\cite{latticegluon,Cucchieri:2009kk} of the gluon propagator
$D(p)$ in LCG.
These numerical data seem to agree with several analytic
predictions \cite{Huber:2015ria,Aguilar:2015nqa,gluonLCG},
e.g.\ the transverse component of $D(p)$ is similar to
the Landau case, with $D(0)$ decreasing when the gauge-fixing
parameter $\xi$ increases.
On the other hand, for the ghost propagator $G(p)$ there is a
wide range of different analytic predictions.
Indeed, the ghost dressing function $p^2\,G(p)$ has been
predicted to be flat (and nonzero) in the infrared limit
\cite{Siringo:2014lva}, or to be suppressed at small momenta
when $\xi$ increases \cite{Huber:2015ria}, or to be null at
$p = 0$ \cite{Aguilar:2015nqa,ghostLCG}.
Numerical results for $G(p)$, however, are not
yet available, since a lattice definition of the Faddeev-Popov
(FP) matrix, corresponding to the minimal LCG on the lattice
\cite{Cucchieri:2009kk}, has not been implemented so far.

In this work we define the FP matrix in lattice minimal
LCG by considering the quadratic expansion of the
corresponding minimizing functional, in analogy with the
Gribov-Zwanziger approach in Landau gauge
\cite{Alkofer:2000wg,Landau1}.
We start by reviewing how the
minimal LCG can be fixed on the lattice, in Sec.\ \ref{sec:mlcg}.
We then consider the quadratic form obtained from the
second variation of the LCG minimizing functional and
its relation to the FP operator in the continuum formulation.
First results for the ghost propagator in LCG are
shown in Sec.\ \ref{sec:ghost} for the SU(2) and SU(3) gauge groups.
Finally, in the last section we present our conclusions.


\section{Minimal Linear Covariant Gauge}
\label{sec:mlcg}

The minimal LCG can be obtained
\cite{Cucchieri:2009kk} by minimizing the functional
\begin{eqnarray}
{\cal E}_{\text{LCG}}[U;\Lambda;h] \,&\equiv&\,
\Re \, \Tr \, \sum_{\vec{x}\in\Lambda_x} \,
\biggl\{ \, \left[ \, i \, h(\vec{x}) \, \Lambda(\vec{x}) \, \right]
\nonumber \\
 &-&\,\sum_{\mu = 1}^d \, \left[ \,
     h(\vec{x}) \, U_{\mu}(\vec{x}) \,
h(\vec{x} + \vec{e}_{\mu})^{\dagger} \, \right] \,
    \biggr\} \; , \mbox{\phantom{ooo}}
\label{eq:minimizingLCG}
\end{eqnarray}
with the remark that, in the numerical minimization, the link
variables $U_{\mu}(\vec{x})$ are gauge-transformed, while the
$\Lambda(\vec{x})$ matrices are not.
The above definition applies to a $d$-dimensional Euclidean
lattice $\Lambda_x$ ---usually with periodic boundary
conditions--- for an SU($N_c$) gauge theory.
Here, $\vec{e}_{\mu}$ is a vector of length $a$ in the
positive $\mu$ direction, $a$ is the lattice spacing, the
vectors $\vec{x}$ have components $x_{\mu} \in \{ a, 2a, \ldots,
Na$\} so that the lattice volume $V$ is equal to $N^d$,
we indicate with $\Tr$ the trace in color space, $\Re$ selects
the real part and $^{\dagger}$ stands for the Hermitian conjugate.
Also, $\{ U_{\mu}(\vec{x}) \} \in$ SU($N_c$) is a given
thermalized link configuration and $\{ h(\vec{x}) \} \in$ SU($N_c$)
is a gauge transformation.
Both the $U_{\mu}(\vec{x})$ and $h(\vec{x})$ matrices are in the
$N_c \times N_c$ (fundamental) representation.
For the $N_c^2 - 1$ traceless Hermitian generators $\lambda^b$ of
SU($N_c$) we use the normalization $ \Tr ( \lambda^b \lambda^c ) = 2
\delta^{bc}$.
Finally, $ \Lambda(\vec{x}) \equiv \sum_{b} \Lambda^b(\vec{x})
\lambda^b$ are (Hermitian) matrices belonging to the SU($N_c$) Lie
algebra and the $\Lambda^b(\vec{x})$ are random real numbers,
usually Gaussian-distributed around zero with a width $\sigma=
\sqrt{\xi}$.

The first and second variations of ${\cal E}_{\text{LCG}}[U;h]$ can
be obtained \cite{Zwanziger:1993dh} by considering for the gauge
transformation a one-parameter subgroup $ h(\tau; \vec{x})
\equiv \exp{ \Bigl[ i \tau \sum_{b} \gamma^b(\vec{x})
\lambda^b \Bigr] }$, where the parameter $\tau$ and the factors
$\gamma^b(\vec{x})$ are real.
Then, by expanding the functional ${\cal E}_{\text{LCG}}[U;\Lambda;h](\tau)$
around a minimum $\{ U_{\mu}(\vec{x}) \}$ up to terms linear
in $\tau$, and by using periodicity, one finds that the stationarity
condition ${\cal E}_{\text{LCG}}[U;\Lambda;h]^{\prime}(0)=0$ ---where
$^{\prime}$ indicates the derivative with respect to the parameter
$\tau$--- gives
\begin{equation}
0 \, = \,
\Re \, \Tr \,\lambda^b \, \biggl[ \, - \,
       \Lambda(\vec{x}) \,  + \, \sum_{\mu = 1}^d \,
      \frac{U_{\mu}(\vec{x}) \, - \, U_{\mu}(\vec{x} - \vec{e}_{\mu})}{i}
          \, \biggr]
\label{eq:Ederiv3}
\end{equation}
for any lattice site $\vec{x}$ and color index $b$.
One usually defines the lattice gauge field $A_{\mu}(\vec{x} + \vec{e}_{\mu}/2)
= \sum_{b} A_{\mu}^{b}(\vec{x} + \vec{e}_{\mu}/2) \, \lambda^b$
through the relation
\begin{eqnarray}
\!\!\!\!
A_{\mu}(\vec{x} + \vec{e}_{\mu}/2) \,&\equiv&\, \frac{1}{2 \,i}\,
   \left[ \, U_{\mu}(\vec{x})
     - U_{\mu}^{\dagger}(\vec{x}) \, \right] \nonumber \\[2mm]
    & & \quad - \, \1 \,
        \frac{\Tr}{2 \,i\,N_c}\, \left[ \, U_{\mu}(\vec{x})
           - U_{\mu}^{\dagger}(\vec{x}) \, \right] \; ,
\mbox{\phantom{o}}
\label{eq:defAgen}
\end{eqnarray}
where $\1$ is the $N_c \times N_c$ identity matrix,
yielding $ A_{\mu}^{b}(\vec{x} + \vec{e}_{\mu}/2) = \Re \Tr \,
\left[ \lambda^{b} U_{\mu}(\vec{x}) / ( 2\,i) \right]$.
Then, if we indicate with
\begin{eqnarray}
\!\!\!\!
\left( \nabla\cdot A^b \right)(\vec{x}) \,&\equiv&\,
  \sum_{\mu = 1}^d \, \left(
    \nabla_{\mu} \, A_{\mu}^{b} \right)(\vec{x}) \nonumber \\[1mm]
&\equiv&\, \sum_{\mu = 1}^d \, A_{\mu}^{b}(\vec{x}
    + \vec{e}_{\mu}/2) -
                  A_{\mu}^{b}(\vec{x} - \vec{e}_{\mu}/2)
\mbox{\phantom{oo}}
\label{eq:diverA}
\end{eqnarray}
the lattice divergence of the gauge field and we use $\nabla_{\mu}$
for the symmetrized lattice derivative, Eq.\ (\ref{eq:Ederiv3}) becomes
\begin{equation}
\left(\nabla\cdot A^b \right)(\vec{x}) \, = \,
           \Lambda^b(\vec{x}) \; .
\label{eq:statLCG}
\end{equation}
We also define $ U_{\mu}(\vec{x}) \equiv \exp{\left[ i a g_0
\hat{A}_{\mu}(\vec{x}+\vec{e}_{\mu}/2) \right]}$,
where $\hat{A}_{\mu}(\vec{x})$ is the continuum gauge field
and $g_0$ is the bare coupling constant.
Thus, in the limit of small $a$, we have that $
A^{b}_{\mu}(\vec{x} + \vec{e}_{\mu}/2) = a g_0
   \hat{A}^{b}_{\mu}(\vec{x}+\vec{e}_{\mu}/2) +
     \mathcal{O}(a^3 \, g_0^3)$
and a similar relation applies to $A_{\mu}^{b}(\vec{x})$.
Note that, compared to the usual generators $\tilde{\lambda}^b$
with normalization $\Tr ( \tilde{\lambda}^b \tilde{\lambda}^c )
= \delta^{bc}/2$, we have $\tilde{\lambda}^b = \lambda^b/2$.
This implies that $ 2\hat{A}^{b}_{\mu}(\vec{x}) \approx
2 A^{b}_{\mu}(\vec{x}) / (a\,g_0)$ is the usual gauge field in
the continuum limit.
Also, in the formal continuum limit, i.e.\ $a \to 0$, $N \to + \infty$
with $L \equiv a \, N$ fixed,  the above equation (\ref{eq:statLCG})
becomes
$a^2 g_0 \sum_{\mu = 1}^d \left[
      \partial_{\mu} \hat{A}_{\mu}^{b}(\vec{x})
   + \mathcal{O}(a^2) \right] =
\Lambda^b(\vec{x})$,
which should be compared\footnote{Here, notation and
conclusions are different from Ref.\ \cite{Cucchieri:2009kk}.}
to the (usual) continuum gauge condition
$ 2\sum_{\mu} \partial_{\mu} \hat{A}_{\mu}^{b}(\vec{x})
= \hat{\Lambda}^b(\vec{x})$, i.e.\ the continuum functions
$\hat{\Lambda}^b(\vec{x})$ satisfy the relation $ a^2 g_0
\hat{\Lambda}^b(\vec{x}) \approx 2 \Lambda^b(\vec{x})$.
Moreover, since the lattice parameter $\beta$ is given by
$2 N_c / (a^{4-d} g_0^2)$ in the $d$-dimensional case, by
setting\footnote{When one considers the usual generators
$\tilde{\lambda}^b$, the relation between the lattice and continuum
gauge parameters is
\begin{equation}
\xi \, = \, \frac{2\,N_c\, \hat{\xi}}{\beta}
\label{eq:xilambdatilde}
\end{equation}.}
\begin{equation}
\xi\, =\, \frac{N_c\, \hat{\xi}}{2 \beta}
\label{eq:xilambda}
\end{equation}
we have that
\begin{equation}
\frac{1}{2\xi} \, \sum_{x, b}
\left[ \, \Lambda^b(x) \, \right]^2 \, = \,
\frac{\beta / (2 N_c)}{2\hat{\xi}} \, \sum_{x, b}
\left[ \, a^2 \, g_0 \, \hat{\Lambda}^b(x) \, \right]^2
\end{equation}
goes to
$(2\hat{\xi})^{-1}
\int d^dx \sum_b \, [ \hat{\Lambda}^b(x) ]^2$
in the formal continuum limit.
Thus, the continuum and lattice widths (of the corresponding
Gaussian distributions) are related through the expression
$\sigma=\hat{\sigma}\, \sqrt{N_c / (2 \beta)}$, i.e.\ for
$N_c = 2,3$ one has $\sigma < \hat{\sigma}$ for typical
values of $\beta$ in the scaling region.

\vskip 3mm
In minimal Landau gauge the FP matrix $\mathcal{M}^{bc}(\vec{x},
\vec{y})$ is obtained from the second-order expansion, with
respect to the parameter $\tau$, of the corresponding minimizing
functional, i.e.\ through the relations\footnote{From now on,
we simplify the notation and we do not indicate explicitly
the lower and upper bounds for the summation indices.}
\begin{eqnarray}
\label{eq:E2}
\frac{{\cal E}_{\text{LG}}[U;h]^{\prime\prime}(0)}{2} \,& = &\,
  \sum_{b,\vec{x}} \,
   \gamma^{b}(\vec{x}) \,
     \left( \mathcal{M} \, \gamma \, \right)^{b}(\vec{x})
 \mbox{\phantom{$\;\;\;\;\;\;$}} \\[1mm]
\left( \mathcal{M} \, \gamma \, \right)^{b}(\vec{x}) \,& = &\,
     \sum_{c,\vec{y}} \,
      \mathcal{M}^{bc}(\vec{x},\vec{y})
          \, \gamma^{c}(\vec{y}) \; ,
\end{eqnarray}
where ${\cal E}_{\text{LG}}[U;h]$ is the Landau-gauge
minimizing functional, given by the second term in the above
Eq.\ (\ref{eq:minimizingLCG}).

On the other hand, one can easily verify that the first
term in Eq.\ (\ref{eq:minimizingLCG}) does not contribute
to this second-order expansion in powers of the parameter
$\tau$.
Indeed, the expression multiplying $\tau^2$ is given by
\begin{equation}
    \Tr \, \sum_{b,c,e,\vec{x}} \, \biggl[ \,
           \gamma^b(\vec{x}) \, \gamma^c(\vec{x}) \,
        f^{bce} \, \lambda^e \,
    \Lambda(\vec{x}) \, \biggr] \; .
\label{eq:zeroterm}
\end{equation}
In the above derivation we made use of the Hermiticity of the
matrices $\lambda^b$, $\Lambda(\vec{x})$, and we
employed the cyclic property of the trace and the
commutation relations $ \left[\lambda^b, \, \lambda^c \right]
\equiv 2 i \sum_{e} f^{bce} \, \lambda^e$, where $f^{bce}$ are
the (real) structure constants of the SU($N_c$)
gauge group.
Let us recall that these structure constants are completely
skew-symmetric in all indices \cite{Cornwell}, since the Lie
algebra of the SU($N_c$) group is simple and compact, and the
generators $\lambda^b$ constitute an orthonormal basis
(through a global rescaling).
Therefore, the expression $\sum_{b,c} \gamma^b(\vec{x})
\gamma^c(\vec{x}) f^{bce}$
in Eq.\ (\ref{eq:zeroterm}) is zero $\forall \,e,\,
\vec{x}$.
As a consequence, the second variation of
${\cal E}_{\text{LCG}}[U;\Lambda;h]$ yields the same
matrix obtained in the Landau case, i.e.\
\cite{Zwanziger:1993dh}
\begin{eqnarray}
\mathcal{M}^{bc}(\vec{x},\vec{y}) \,& \equiv &\,
  \sum_{\mu} \, \Biggl\{ \, \Gamma^{bc}_{\mu}(\vec{x}) \,
         \Bigl[ \, \delta_{\vec{x},\,\vec{y}} \, - \,
 \delta_{\vec{x}+\vec{e}_{\mu},\,\vec{y}} \, \Bigr]
            \nonumber \\[0mm]
 & & \quad + \, \Gamma^{bc}_{\mu}(\vec{x}-\vec{e}_{\mu})
   \, \Bigl[ \, \delta_{\vec{x},\,\vec{y}} \, - \,
   \delta_{\vec{x}-\vec{e}_{\mu},\,\vec{y}} \, \Bigr]
     \nonumber \\[2mm]
& & \; - \, \sum_{e} \,
   f^{bec} \Bigl[ \, A^e_{\mu}(\vec{x}-\vec{e}_{\mu}/2) \,
    \delta_{\vec{x}-\vec{e}_{\mu},\,\vec{y}}
      \nonumber \\[-1mm]
 & & \qquad \, - \,
     A^e_{\mu}(\vec{x}+\vec{e}_{\mu}/2)
      \, \delta_{\vec{x}+\vec{e}_{\mu},\,\vec{y}}
              \, \Bigr] \, \Biggr\}
\mbox{\phantom{ooo}}
\label{eq:Mbcxy}
\end{eqnarray}
with
\begin{equation}
\Gamma^{bc}_{\mu}(\vec{x}) \, \equiv \, \Tr \, \left[ \,
   \frac{\lambda^b \, \lambda^c \, + \, \lambda^c \,
          \lambda^b}{4} \, \frac{U_{\mu}(\vec{x}) \, + \,
          U^{\dagger}_{\mu}(\vec{x})}{2} \,
    \right] \; .
\label{eq:defGamma}
\end{equation}
It is immediate to verify that $\mathcal{M}^{bc}(\vec{x},
\vec{y})$ is symmetric under the simultaneous exchanges
$b \leftrightarrow c$ and $\vec{x} \leftrightarrow \vec{y}$.

One can also set
\begin{eqnarray}
\mathcal{M} \,&=&\,
\frac{1}{2}\,\left(\,
\mathcal{M}_{+} \, + \, \mathcal{M}_{-} \, \right) \; ,
\label{eq:MMM} \\[2mm]
\left( \mathcal{M}_{\pm}\;\!\gamma \right)^{b}\!(\vec{x}) \,&
\equiv& \, \left( \mathcal{M}\;\!\gamma \right)^{b}\!(\vec{x})
\pm (\Delta \mathcal{M}\,\gamma)^{b}(\vec{x})
\mbox{\phantom{oooo}}
\label{eq:Mpm} \\[2mm]
(\Delta \mathcal{M})^{bc}(\vec{x},\vec{y}) \,&\equiv&\, \sum_{e}
\, f^{bec} \, \left( \nabla\cdot A^e \right)(\vec{x})\,
\delta_{\vec{x},\vec{y}} \; .
\label{eq:E2symm-bis}
\end{eqnarray}
At the same time, we define the lattice gauge-covariant derivative
by the relation \cite{Zwanziger:1993dh}
\begin{eqnarray}
D_{\mu}^{bc}(\vec{x},\vec{y}) \,&\equiv&\,
  \Gamma^{bc}_{\mu}(\vec{x}) \,
  \Bigl[ \, \delta_{\vec{x}+\vec{e}_{\mu},\vec{y}}
  \, - \, \delta_{\vec{x},\vec{y}} \, \Bigr] \nonumber \\[2mm]
- \, \sum_{e} & &\!\!\!\!\!\!\! f^{bec} \,
   A^e_{\mu}(\vec{x}+\vec{e}_{\mu}/2) \,
\Bigl[ \, \delta_{\vec{x}+\vec{e}_{\mu},\vec{y}}
  \, + \, \delta_{\vec{x},\vec{y}} \, \Bigr] \; .
\mbox{\phantom{ooo}}
\label{eq:Dbcxy}
\end{eqnarray}
Indeed, in the formal continuum limit, we have
$\Gamma^{bc}_{\mu}(\vec{x}) \to \delta^{bc} +
\mathcal{O}(a^2 g_0^2)$, giving
\begin{equation}
\left(D_{\mu}\,\gamma \right)^{b}(\vec{x}) \, \to \,
a \, \left[ \left(D_{\mu}[\hat{A}] \, \gamma\right)^{b}\!(\vec{x}) \,
     + \, \mathcal{O}(a,\,a \, g_0) \right] \; ,
\end{equation}
where $D_{\mu}^{bc}[\hat{A}] \equiv \delta^{bc} \partial_{\mu}
+ 2 g_0 \sum_{e} f^{bce} \hat{A}^e_{\mu}(\vec{x})$
is the continuum gauge-covariant derivative.
[As explained above, with our notation, the continuum gauge field
is given by $2\,\hat{A}^e_{\mu}(\vec{x})$.]
Then, it is easy to verify that
\begin{eqnarray}
\left( \mathcal{M}_{+} \, \gamma \, \right)^{b}(\vec{x})
\,& = &\, - \, \sum_{\mu}\,
   \left[ \, \left( D_{\mu} \, \gamma \right)^b(\vec{x})
  - \left( D_{\mu} \, \gamma \right)^b(\vec{x}-\vec{e}_{\mu})
      \, \right] \nonumber \\[1mm]
 & \equiv &\, - \, \sum_{\mu} \, \left[ \,
 \nabla_{\mu}^{(-)} \left( \, D_{\mu} \, \gamma \, \right)
      \, \right]^b(\vec{x}) \; ,
\label{eq:M+}
\end{eqnarray}
where $\nabla_{\mu}^{(-)}$ is the usual backward lattice
derivative.
Thus, $\mathcal{M}_{+}$ is a lattice discretization of the
continuum operator $ \hat{\mathcal{M}}^{bc}_{+}[\hat{A}]
\equiv - \sum_{\mu} \partial_{\mu} \, D_{\mu}^{bc}[\hat{A}]$
and we have $ \mathcal{M}_{+}^{bc} = a^2 \left[
\hat{\mathcal{M}}^{bc}_{+}[\hat{A}] + \mathcal{O}(a,a g_0)
\right]$ in the limit $a \to 0$.
Also, from the above Eq.\ (\ref{eq:Dbcxy}) we can define the
transpose lattice gauge-covariant derivative
\begin{eqnarray}
\left(D_{\mu}^T\right)^{bc}(\vec{x},\vec{y}) \,&\equiv&\,
\Gamma^{bc}_{\mu}(\vec{x}-\vec{e}_{\mu})\,
\delta_{\vec{x}-\vec{e}_{\mu},\vec{y}} \, - \,
\Gamma^{bc}_{\mu}(\vec{x}) \,
   \delta_{\vec{x},\vec{y}}\nonumber \\[2mm]
& & \, + \, \sum_{c,e} \,
   f^{bec} \,\Bigl[ \,A^e_{\mu}(\vec{x}+\vec{e}_{\mu}/2) \,
   \delta_{\vec{x},\vec{y}}  \nonumber  \\[1mm]
& & \qquad + \,
     A^e_{\mu}(\vec{x}-\vec{e}_{\mu}/2) \,
     \delta_{\vec{x}-\vec{e}_{\mu},\vec{y}} \, \Bigr] \; ,
\end{eqnarray}
which goes to $- a \left[ D_{\mu}^{bc}[\hat{A}]
+ \mathcal{O}(a,\,a \, g_0,\,a\,g_0^2) \right]$ in the
formal continuum limit.
Then, one can verify that
\begin{equation}
\left( \mathcal{M}_{-} \, \gamma \, \right)^{b}(\vec{x}) \,=\,
 \sum_{\mu} \, \left[ \,
      D_{\mu}^{T} \, \left( \nabla_{\mu}^{(+)} \, \gamma \,
           \right)\, \right]^b(\vec{x}) \; ,
\label{eq:M-D}
\end{equation}
where $\nabla_{\mu}^{(+)}$ is the usual forward lattice derivative,
and we can identify $\mathcal{M}_{-}$ with a lattice discretization
of the continuum operator $ \hat{\mathcal{M}}^{bc}_{-}[\hat{A}] \equiv
- \sum_{\mu} D_{\mu}^{bc}[\hat{A}]\,\partial_{\mu}$.
Indeed, in the limit $a \to 0$, we have that $\mathcal{M}_{-}$
goes to $a^2 \left[ \hat{\mathcal{M}}^{bc}_{-}[\hat{A}] +
\mathcal{O}(a,a g_0,a g_0^2) \right]$.
Finally, since the transpose of the backward lattice derivative
$\nabla_{\mu}^{(-)}$ is given by $- \nabla_{\mu}^{(+)}$, it is
evident that $\mathcal{M}_{-}^{T} = \mathcal{M}_{+}$ [and thus
$\mathcal{M}_{+}^{T} = \mathcal{M}_{-}$].
Therefore, the matrix $\mathcal{M}$ in Eq.\ (\ref{eq:MMM}) can be written
as $\,\left(\mathcal{M}_{+} + \mathcal{M}_{+}^{T}
\right) / 2 = \left(\mathcal{M}_{-}^{T} + \mathcal{M}_{-}
\right) / 2$,
which is clearly symmetric (and real), in agreement with
the expression (\ref{eq:Mbcxy}).

\vskip 3mm
One should recall that, in the Landau case, the expression
(\ref{eq:E2symm-bis}) is trivially null, due to the
transversality condition $\left( \nabla\cdot A^e \right)
(\vec{x}) = 0$, and one has that (on the lattice as well as in
the continuum) $\mathcal{M} = \mathcal{M}_{+} =
\mathcal{M}_{-}$.
This is not the case in LCG: the matrices $\mathcal{M},
\mathcal{M}_{+}$ and $\mathcal{M}_{-}$ are different.
However, since the expression (\ref{eq:E2symm-bis}) for
$(\Delta \mathcal{M})^{bc}(\vec{x},\vec{y})$ is
skew-symmetric under the simultaneous exchanges
$b \leftrightarrow c$ and $\vec{x} \leftrightarrow \vec{y}$,
these matrices cannot be distinguished as quadratic forms.
This is a general result: given a square matrix $M$, the
corresponding quadratic form depends \cite{Abadir-Magnus}
only on its symmetric part $(M + M^{T})/2$.
Thus, the FP matrix obtained from the second variation of a
minimizing functional is defined modulo an arbitrary,
additive skew-symmetric term.
The situation is similar to the problem of defining a
conserved energy-momentum tensor $T^{\mu\nu}$ in field
theory \cite{Ramond-etc}, where the condition $\partial_{\mu}
T^{\mu\nu} = 0$ implies that $T^{\mu\nu}$ is defined modulo
an additive term $\partial_{\rho} f^{\mu \nu \rho}$, with
$f^{\mu \nu \rho} = - f^{\rho \nu \mu}$.
This freedom is related to the freedom of adding to the
Lagrangian a null (surface) divergence term and it is
usually employed to make the energy-momentum tensor
symmetric and gauge invariant (in the case of a gauge
theory).
In our case we can use the freedom of adding to the symmetric
lattice FP matrix $\mathcal{M}$ the skew-symmetric term
$\Delta\mathcal{M}$ in order to
obtain the lattice FP matrix $\mathcal{M}_{+} =
- \nabla^{(-)} \cdot D$, thus getting (in the limit $a \to 0$)
the usual continuum result $ - \sum_{\mu} \partial_{\mu} \,
D_{\mu}^{bc}[\hat{A}]$.
Equivalently, we could add to the minimizing
functional ${\cal E}_{\text{LCG}}[U;\Lambda;h]$
the null term $ - \Re \Tr \sum_{\vec{x}} i \left[ h(\vec{x}),
\Lambda(\vec{x}) \right] h(\vec{x})^{\dagger}$, which obviously
does not affect the minimizing procedure.
Indeed, by considering the one-parameter subgroup $h(\tau; \vec{x})$
and by expanding the above expression at order $\tau^2$ we find
---by a convenient reordering of the null terms and by
using the stationarity condition (\ref{eq:statLCG})---
the quadratic expression
$\sum_{b,c,\vec{x},\vec{y}}
 \gamma^{b}(\vec{x}) (\Delta \mathcal{M})^{bc}(\vec{x},\vec{y})
 \gamma^{c}(\vec{y})$.

Let us note that, following the usual continuum FP approach,
the matrix $\mathcal{M}_{+}$ can also be obtained
from a variation of the gauge condition (\ref{eq:Ederiv3})
with respect to the gauge transformation $ h(\vec{x})
\equiv \exp{ \Bigl[ i \sum_{b} \gamma^b(\vec{x})
\lambda^b \Bigr] }$, namely by evaluating the
functional derivative of
\begin{eqnarray}
& & \Re \, \Tr \,\lambda^b \, \biggl[ \, - \,
       \Lambda(\vec{x}) \,  + \, i \, \sum_{\mu = 1}^d \,
 h(\vec{x} - \vec{e}_{\mu}) \,
  U_{\mu}(\vec{x} - \vec{e}_{\mu}) \,
  h(\vec{x})^{\dagger} \nonumber \\
& & \quad - \, h(\vec{x}) \,U_{\mu}(\vec{x}) \,
  h(\vec{x} + \vec{e}_{\mu})^{\dagger}
          \, \biggr]
\end{eqnarray}
with respect to $\gamma^c(\vec{y})$.
This adds a heuristic motivation for the consideration of $\mathcal{M}_+$
among all the possible discretizations of the continuum FP matrix in LCG.


\section{The Ghost Propagator}
\label{sec:ghost}

In order to evaluate the ghost propagator
\begin{equation}
G(k) \, \equiv \, \frac{1}{V}\,
\sum_{b,\vec{x},\vec{y}}\,
e^{i \vec{k}\cdot\vec{x}} \,
\left[\left(\mathcal{M}_{+}\right)^{-1}\right]^{bb}(\vec{x},\vec{y})
\, e^{-i \vec{k}\cdot\vec{y}}
\label{eq:gk}
\end{equation}
in LCG we need to invert the FP matrix $\mathcal{M}_{+}$, defined in
Eqs.\ (\ref{eq:Mbcxy}), (\ref{eq:Mpm}) and (\ref{eq:E2symm-bis})
above.
Since this matrix is real but not symmetric, its eigenvalues and
eigenvectors do not need to be real and the nonreal eigenvalues and
eigenvectors occur in complex-conjugate pairs.
Also, one can easily check that only the symmetric (respectively,
skew-symmetric) part of the inverse matrix
$\left(\mathcal{M}_{+}\right)^{-1}$ contributes to the
real (respectively, imaginary) part of the r.h.s.\ of Eq.\
(\ref{eq:gk}).
Thus, the ghost propagator in LCG is in general a complex quantity,
while in Landau gauge it is always real.
Finally, in order to invert $\mathcal{M}_{+}$ one cannot use, as
in the Landau cause, the conjugate gradient method (since
$\mathcal{M}_{+}$ is not symmetric), i.e.\ one needs a more general
iterative Krylov subspace method, applicable to generic non-singular
matrices \cite{Saad}.

\begin{figure}[t]
\centering
\includegraphics[scale=0.65]{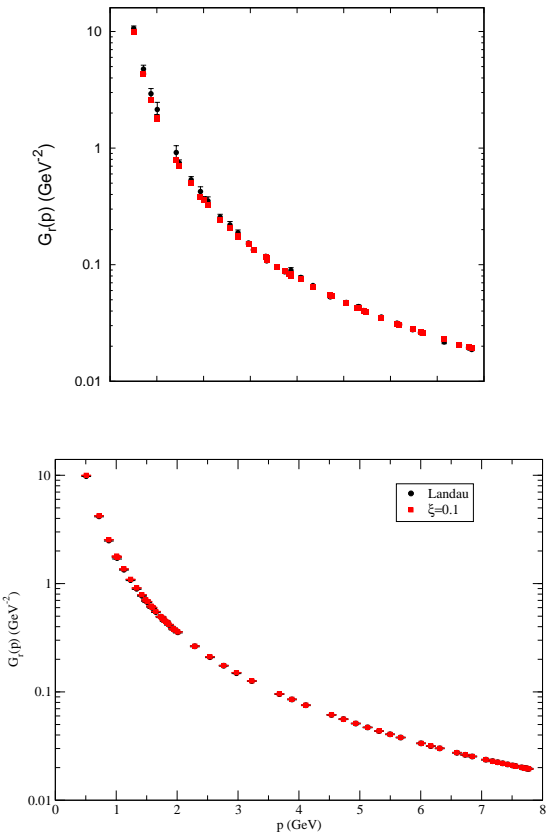}
\
\vskip 4mm
\includegraphics[scale=0.3]{ghost24.eps}
\caption{\label{fig} The (real part of the) ghost propagator
    $G_{\text{r}}(p)$ in minimal
    LCG (\protect\marksymbol{square*}{red}) and in Landau gauge
    (\protect\marksymbol{*}{black}), as a function of the lattice
    momentum $p$, with $p_{\mu}(k)
    = 2 \sin{(\pi k_{\mu}/N)}$ and $k_{\mu} = 1,2,\ldots,N/2$.
    Note the logarithmic scale on the $y$ axis.
    Both $G_{\text{r}}(p)$ and $p$ are in physical units.
    {\bf Top}: SU(2) case with $V = 24^4$, $\beta = 2.4469$ and
    $\xi = 0.1$, corresponding to the continuum value $\hat{\xi}
    = 0.24469$, for 60 thermalized configurations.
    {\bf Bottom}: SU(3) case with $V = 24^4$, $\beta = 6.0$ and
    $\xi = \hat{\xi} = 0.1$ for 79 thermalized configurations.
}
\end{figure}

We have performed tests evaluating the ghost
propagator in the four-dimensional case, for the SU(2) and
SU(3) gauge groups, respectively using the bi-conjugate
gradient stabilized algorithm and the generalized conjugate
residual for the inversion of the FP matrix \cite{Saad}.
For these two gauge groups we have considered lattice
couplings $\beta = 2.4469$ and $\beta = 6.0$, respectively,
which both correspond \cite{Cucchieri:2007zm} to lattice
spacing $a \approx 0.102\; \mbox{fm}$.
Simulations have been done for lattice volumes $V = 16^4$
and $24^4$.
Thus, for the larger volume, the (nonzero) lattice momenta
range from about 500 MeV to about 7.7 GeV.
For each thermalized gauge configuration we have generated
20 sets of Gaussian-distributed $\{ \Lambda(x) \}$ matrices,
with variance\footnote{In the SU(2) case we used the $\lambda^b$
generators with normalization $ \Tr ( \lambda^b \lambda^c )
= 2 \delta^{bc}$.
For SU(3) we employed the $\tilde{\lambda}^b = \lambda^b/2$
generators with normalization $\Tr ( \tilde{\lambda}^b \tilde{\lambda}^c )
= \delta^{bc}/2$.
Thus, in the former case we have
$\hat{\xi} = \beta \, \xi$ [see Eq.\ (\ref{eq:xilambda})],
while in the latter we find $\hat{\xi} = \beta \, \xi / 6$
[see Eq.\ (\ref{eq:xilambdatilde})].}
$\xi = 0.1$ in the SU(2) case (corresponding to
$\hat{\xi} = 0.24469$), and $\xi = \hat{\xi} = 0.1, 0.2$ and 0.3
in the SU(3) case.\footnote{For $V=24^4$, in the SU(3) case,
simulations have been done only for $\xi = \hat{\xi} = 0.1$.
These are the data shown in the bottom plot of Fig.\ \ref{fig}.
(Results for the $V=16^4$ cases are similar.)}
The ghost propagator has been evaluated using a point source
for the inversion \cite{Boucaud:2005gg}.
Results are reported in Fig.\ \ref{fig}, where we
compare the real part of the ghost propagator $G_{\text{r}}(p)$
in minimal LCG with the corresponding data in Landau gauge, using
the same set of thermalized configurations.
Clearly the data in LCG are in agreement, within error
bars, with the data in Landau gauge.
Let us mention that, in continuum analytic works, one usually finds that
$G(p)$ is real \cite{Huber:2015ria,Aguilar:2015nqa,ghostLCG,Davydychev:1996pb}.
A numerical check of this result is postponed to a future study.

\section{Conclusions}
\label{sec:conclusions}

In this work we have discussed the relation among the FP
matrix in lattice minimal LCG and the second variation of
the corresponding minimizing functional, following the
usual Gribov-Zwanziger approach for Landau gauge
\cite{Alkofer:2000wg,Landau1}.
In particular, we have chosen the matrix $\mathcal{M}_{+}$
[see Eqs.\ (\ref{eq:Mbcxy}), (\ref{eq:Mpm}) and (\ref{eq:E2symm-bis}) above]
as a natural lattice
discretization of the LCG continuum FP operator $ - \sum_{\mu}
\partial_{\mu} D_{\mu}^{bc}[\hat{A}]$.
We have also carried out some tests for the numerical inversion
of the matrix $\mathcal{M}_{+}$ and evaluated the ghost
propagator.
Preliminary results for the (real part of the) ghost
propagator $G_{\text{r}}(p)$ show no detectable difference
with the corresponding lattice data in Landau gauge.
Of course, numerical simulations for larger physical
volumes, different lattice spacings $a$ and gauge
parameters $\xi$ should be performed before
any final conclusion is drawn about the behavior in
minimal LCG of $G_{\text{r}}(p)$ at small momenta.
One should also recall that, in the continuum,
there are different possible setups for the ghost
sector in LCG (see e.g.\ Appendix A in Ref.\
\cite{Alkofer:2000wg}).
The FP matrix $\mathcal{M}_{+}$, considered here,
corresponds to the usual choice of complex ghost/antighost
fields, without enforcing the ghost-antighost symmetry, which
is naturally realized in Landau gauge.
On the other hand, for a generic linear covariant gauge with
$\xi\neq0$, this choice is at odds with demanding Hermiticity
of the underlying Lagrangian, which requires in principle
the introduction of a doublet of real ghost/antighost fields
\cite{Alkofer:2000wg,Kugo:1979gm}.
Clearly, it would be important to analyze if and how the other setups
can also be implemented on the lattice in minimal LCG.
Another open question is how to define an appropriate Gribov region,
similarly to the Landau-gauge case.
A more detailed analysis of these issues
will be presented elsewhere.


\section*{Acknowledgments}

A.C.~and T.~M.~acknowledge partial support from CNPq.
A.C.\ also acknowledges partial support from FAPESP (grant
$\#$ 16/22732-1). The research of D.D.~and M.R.~is supported by KU
Leuven IF project C14/16/067. The authors 
O.O.~and P.J.S.~acknowledge the Laboratory for Advanced Computing at 
University of Coimbra (\url{http://www.uc.pt/lca}) for providing access 
to the HPC computing resource {\tt Navigator}.
P.J.S. acknowledges support by FCT under contracts SFRH/BPD/40998/2007
and SFRH/BPD/109971/2015.
The SU(3) simulations were done using {\tt Chroma} \cite{Edwards:2004sx}
and {\tt PFFT}~\cite{Pippig2013} libraries.



\end{document}